\documentclass{article}
\title{Protocol for making a $2$-qutrit entangling gate in the Kauffman-Jones version of $SU(2)_4$}
\author{Claire Levaillant}
\usepackage{amsmath, amsthm}
\usepackage{epsfig}
\newcommand{\nts}{\negthickspace}
\begin{document}
\maketitle
\begin{center}
\textbf{Acknowledgement}
\end{center}
This work was done with Michael Freedman. The author thanks him for his quite nice and fruitful mentoring. She thanks Stephen Bigelow for his comments and help with a better understanding of interferometry.

\begin{center}
\textbf{Abstract}\end{center}
The following paper provides a protocol to physically generate a $2$-qutrit entangling gate in the Kauffman-Jones version of $SU(2)$ Chern-Simons theory at level $4$. The protocol uses elementary operations on anyons consisting of braids, interferometric measurements, fusions and unfusions and ancilla pair creations. 
%Other applications include universal quantum computation and making the Hadamard gate
%$$H=\frac{1}{\sqrt{2}}\begin{pmatrix}1&1\\1&-1\end{pmatrix}$$

\section{Introduction}
\newtheorem{Theorem}{Theorem}
\indent The Brylinski couple, in a work \cite{BRY} dating from $2001$, proves that universal single qudit gates together with any $2$-qudit entangling gate are sufficient for universal quantum computation. For qubits, there is an independent proof of this fact by \cite{BB2}. The goal of our present paper is to provide a protocol for producing a $2$-qutrit entangling gate in the framework of the Kauffman-Jones version \cite{KL} of $SU(2)$ Chern-Simons theory at level $4$. Throughout the paper, we assume that the reader is familiar with this theory and a nice reference for it is \cite{KL}. We also assume that the reader is familiar with interferometric measurements and we refer the reader to \cite{BO}, \cite{BO2}, \cite{BO3}.

The anyonic system which we consider is the qutrit formed by four anyons of topological charge $2$. The action by the braid group on this qutrit as well as a pair fusion action due to Michael Freedman on the same qutrit are extensively studied in \cite{BL} and \cite{CL2} respectively. The core ideas to produce an entangling gate on the $2$-qutrit are to introduce an ancilla which plays the role of a mediator between the left and right qutrits and use it to entangle both qutrits before separating the two qutrits again. Along the way, we need a qubit swap braid that was introduced in \cite{CL}. Our protocol uses braids, interferometric measurements, fusions and unfusions of anyons, vacuum pair creation and recovery procedures. These ideas originate in \cite{MO}. 

\section{A $2$-qutrit entangling gate}
\begin{Theorem}
The following protocol
\begin{center}
\epsfig{file=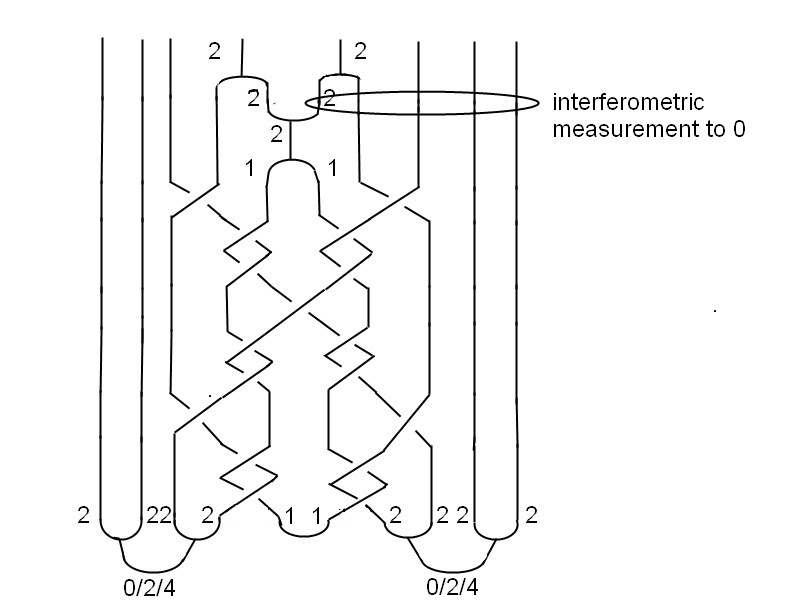, height=12cm}
\end{center}
produces the following entangling gate
$$\begin{array}{l}\\\begin{array}{cc}&\begin{array}{l}|00\nts>|02\nts>|04\nts>|20\nts>|22\nts>|24\nts>|40\nts>|42\nts>|44\nts>
\end{array}\\&\\
\begin{array}{l}|00>\\\\|02>\\\\|04>\\\\|20>\\\\|22>\\\\|24>\\\\|40>\\\\|42>\\\\|44>\end{array}&\begin{pmatrix}
e^{\frac{i\,2\pi}{3}}&&&&&&&&\\
&&&&&&&&\\
&1&&&&&&&\\
&&&&&&&&\\
&&e^{\frac{i\,2\pi}{3}}&&&&&&\\
&&&&&&&&\\
&&&1&&&&&\\
&&&&&&&&\\
&&&&e^{\frac{i\,2\pi}{3}}&&&&\\
&&&&&&&&\\
&&&&&1&&&\\
&&&&&&&&\\
&&&&&&e^{\frac{i\,2\pi}{3}}&&\\
&&&&&&&&\\
&&&&&&&1&\\
&&&&&&&&\\
&&&&&&&&e^{\frac{i\,2\pi}{3}}
\end{pmatrix}\end{array}\\\\\end{array}$$

\noindent In the protocol leading to the gate above, there are two measurements. One is a fusion measurement and the other one is an interferometric measurement. In each case, if the measurement outcome is unlike on the drawing, a recovery procedure is needed and is provided along the proof of the Theorem.

\end{Theorem}

\noindent\textsc{Proof of the Theorem.}
Before we start the discussion, we will convince the reader that the announced gate is indeed \textbf{entangling}. For instance, it is visible that it entangles the vector $v=|00> +|02>+|04>+|20>+|22>+|24>$. Indeed, if $G$ denotes the gate, $G$ acts on this vector by
$$Gv=e^{\frac{2i\pi}{3}}(|00>+|04>+|22>)+(|02>+|20>+|24>)$$
Suppose we could write
$$Gv=(a_0|0>+a_2|2>)\otimes(b_0|0>+b_2|2>+b_4|4>)$$
Then, it would follow that
$$\left\lbrace\begin{array}{ccccccc}
a_0b_0&=&a_0b_4&=&a_2b_2&=&e^{\frac{i2\pi}{3}}\\
a_0b_2&=&a_2b_0&=&a_2b_4&=&1
\end{array}\right.$$
This is impossible.

By choice, we won't detail the challenging computations and will rather outline how we proceed. Below is a summary of how the protocol is built.
\begin{itemize}
\item
We use an ancilla to entangle both qutrits.\\
\item Once the ancilla has approached the two qutrits, it acts on them. \\
\item Using the information the ancilla got from the qutrits, it entangles them. \\
\item Now comes the qutrits separation phase. The two qutrits have to get back into shape and the ancilla must vanish, leaving the qutrits separated but entangled. \\
\item Either it is arbitrarily decided that the ancilla did a good job and vanished properly or it is arbitrarily decided that the ancilla messed up with the two qutrits. In the latter case, a recovery is needed.
\end{itemize}
$$\begin{array}{l}\end{array}$$

\underline{Now in more mathematical details}.\\\\ First, we bring a pair of $1$'s out of the vacuum and we do a full twist on $(2211)_0$ and $(1122)_0$ on each side, which results in introducing a $2$ charge line between each input and the ancilla by \cite{CL}, $\S\,5$.
\begin{center}
\epsfig{file=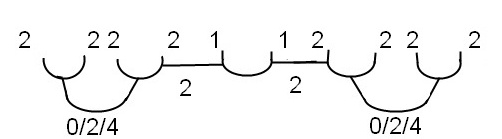, height=3cm}
\end{center}

We then use these charge lines to do the braid
\begin{center}
\epsfig{file=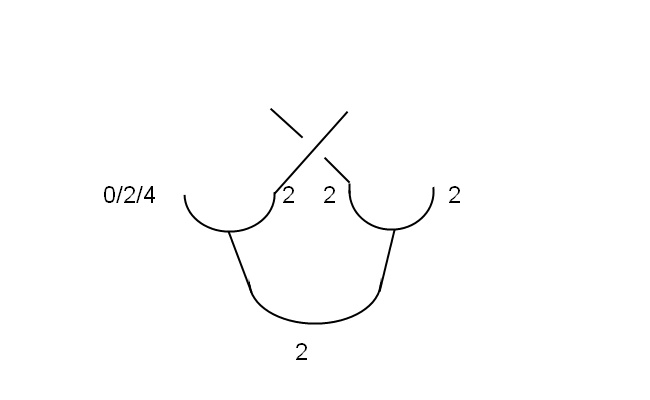, height=6cm}
\end{center}
with the vertical mirror image for the picture relating to the right qutrit. All we do on the left input side, we do the same on the right input side. The protocol is entirely symmetric, thus we will only take care of describing what happens on the left input side. The reader should keep in mind that the same actions are being performed simultaneously on the right input side. A $\sigma_2$-braid on $(0222)_2$ or $(4222)_2$ is simply a multiplication by a phase while by \cite{BL} a $\sigma_2$-braid on $(2222)_2$ results in a superposition of $0$ and $4$. We then transmit this information from the qutrit to the center by doing another full twist like on the drawing. When the qutrit input is $0$ or $4$, the full twist on $(2211)_2$ results in a new qubit swap (cf \cite{CL}, $\S\,5$), while when the qutrit input is a $2$, a full twist on $(0211)_2$ or $(4211)_2$ simply results in a multiplication by a phase since for instance by the fusion rules $2$ and $0$ or $2$ and $4$ can only fuse into $2$.
Now braiding anyons $5$ and $6$ will be likely to entangle the left and right qutrits. Note that in the case when the input is $|22>$, the "bottom part" of the ancilla pair of $1$'s carries a superposition of charges $1$ and $3$ after the entangling braid is done. The work is not yet over.
First, we must get back to our initial qutrits $2222$. Second, we must destroy the central part which contains the ancilla pair. By doing another full twist, we put the label of the edge adjacent to the ancilla pair and to the input to $2$ \textbf{independently from the input}. This step is very important as it allows the outcome of the forthcoming fusion measurement to be completely independent from the $2$-qutrit input. Next, we do a single braid in order to retrieve the initial shape of the input using the fact that a full twist on $(2222)_2$ is a multiplication by a phase by \cite{CL} or \cite{CL2}. It remains to dispose of the ancilla and separate the left and right qutrits. To the first aim, we do the simplest thing consisting of fusing anyons $5$ and $6$. However, when doing this fusion measurement, the outcome can either be $2$ which leads to the gate of the Theorem or $0$, leading in the latter case to a different gate, namely to

$$\begin{array}{l}\\\begin{array}{cc}&\begin{array}{l}|00\nts>|02\nts>|04\nts>|20\nts>|22\nts>
|24\nts>|40\nts>|42\nts>|44\nts>
\end{array}\\&\\
\begin{array}{l}|00>\\\\|02>\\\\|04>\\\\|20>\\\\|22>\\\\|24>\\\\|40>\\\\|42>\\\\|44>\end{array}&\begin{pmatrix}
1&&&&&&&&\\
&&&&&&&&\\
&e^{-\frac{2i\pi}{3}}&&&&&&&\\
&&&&&&&&\\
&&1&&&&&&\\
&&&&&&&&\\
&&&e^{-\frac{2i\pi}{3}}&&&&&\\
&&&&&&&&\\
&&&&e^{-\frac{2i\pi}{3}}&&&&\\
&&&&&&&&\\
&&&&&e^{-\frac{2i\pi}{3}}&&&\\
&&&&&&&&\\
&&&&&&1&&\\
&&&&&&&&\\
&&&&&&&e^{-\frac{2i\pi}{3}}&\\
&&&&&&&&\\
&&&&&&&&1
\end{pmatrix}\end{array}\\\\\end{array}$$

Fortunately, there exists a recovery procedure which allows to always produce the same gate (as stated in the Theorem). Before we start talking about the recovery, we describe the end of the protocol.
Suppose we have what we decided to be the winning fusion outcome, that is a $2$. By using a pair of $2$'s and fusing one anyon of the pair into $2$, we unfuse the $2$ into two $2$'s (this could take several tries, for instance, if the $2$ unfused into a $2$ and a $4$, we would fuse the $2$ and the $4$ and start over again). Then we run an interferometric measurement on the five anyons $6$, $7$, $8$, $9$, $10$ and hope to measure $0$. If so, it remains to fuse anyons $4$ and $5$ on one hand and $6$ and $7$ on the other hand to be back to a $2$-qutrit. If the interferometric measurement outcome is rather $4$, it will suffice to fuse a pair of $4$'s into anyons $5$ and $6$, an operation called FFO after Michael Freedman in \cite{CL2} and finish with the two fusions like previously. Finally, the outcome cannot be $2$, because specific to this $SU(2)$ Chern-Simons theory at level $4$, the $6j$-symbol with only $2$'s is zero.
$$\left\lbrace\begin{array}{ccc}2&2&2\\2&2&2\end{array}\right\rbrace=0$$

The recovery procedure uses $\sigma_1$-braids. We provide below the R-matrix squared on two anyons of topological charge $2$.
$$R(2,2)^2=\begin{pmatrix} 1&&\\&e^{-\frac{i2\pi}{3}}&\\&&1\end{pmatrix}$$
Suppose the fusion measurement outcome is $0$. Bring a pair of $2$'s out of the vacuum and run the same interferometric measurement as before on anyons $6,7,8,9,10$, then follow the exact same procedure as already described in order to separate the two qutrits, except now the measurement outcome could be $2$. In the latter case, we braid anyons $5$ and $6$ before remeasuring anyons $6$, $7$, $8$, $9$ and iterate the process until we eventually measure $0$. Obtain the gate written above instead of the one from the Theorem. Now the recovery itself. Do a full twist inverse $\sigma_1$-braid on each input like on the drawing below.
\begin{center}
\epsfig{file=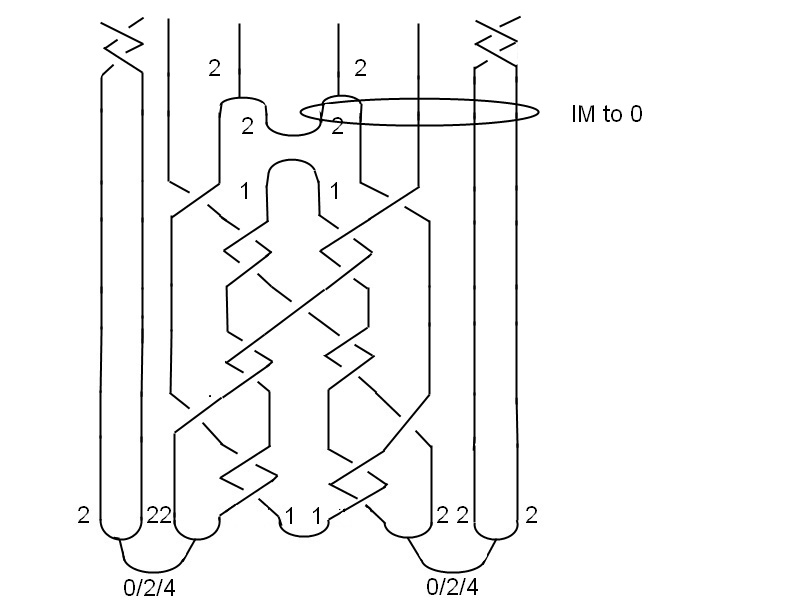, height=12cm}
\end{center}
$$\begin{array}{l}\end{array}$$
Get the new matrix with $1$'s on the diagonal, except in diagonal position $|22>$ where the coefficient reads the phase $e^{\frac{i2\pi}{3}}$.\\
From there, it is visible on the matrices that the following recovery procedure will lead to the gate from the Theorem.\\
\begin{center} \textit{Algorithm}\end{center}
Start the protocol from the beginning again. At the level of the fusion measurement,
\begin{itemize}
\item If we measure a $0$ again, then we obtain the gate from Theorem $2$.
\item If we measure a $2$ the second time, do the same recovery as before with the $\sigma_1$-braids, which this time leads to the identity matrix. Then start all over again.
\end{itemize}

\end{document}